\documentclass{ws-procs9x6}

\begin{document}

\title{The INAF ASTRI Project in the framework of CTA}

\author{N. La Palombara$^*$, P. Caraveo, M. Fiorini, L. Stringhetti}

\address{INAF -- IASF Milano, Via E. Bassini 15, 20133 Milano (I)\\
$^*$E-mail: nicola@iasf-milano.inaf.it}

\author{R. Canestrari, R. Millul, G. Pareschi}

\address{INAF - Osservatorio Astronomico di Brera, via Brera 28, 20121 Milano (I)}

\author{O. Catalano, M. C. Maccarone, S. Vercellone}

\address{INAF -- IASF Palermo, Via U. La Malfa 153, I-90146 Palermo (I)}

\author{E. Giro}

\address{INAF - Osservatorio Astronomico di Padova, Vicolo Osservatorio 5, 35122 Padova (I)}

\author{G. Tosti}

\address{Universit\'{a} di Perugia, Dip. Fisica, Via A. Pascoli, I-06123 Perugia (I)}

\author{for the ASTRI Collaboration (http://www.brera.inaf.it/astri)}

\begin{abstract}
The ASTRI project aims to develop, in the framework of the Cherenkov Telescope Array, an end-to-end prototype of the small-size telescope, devoted to the investigation of the energy range $\sim$ 1--100 TeV. The proposed design is characterized by two challenging but innovative technological solutions which will be adopted for the first time on a Cherenkov telescope: a dual-mirror Schwarzschild--Couder configuration and a modular, light and compact camera based on Silicon photo-multipliers. Here we describe the prototype design, the expected performance and the possibility to realize a mini array composed by a few such telescopes, which shall be placed at the final CTA Southern Site.
\end{abstract}


\bodymatter

\section{CTA and the ASTRI project}\label{sec:project}

The Cherenkov Telescope Array (CTA, Acharya et al. 2013\cite{Acharya+13}) is the project of a new array of several Imaging Atmospheric Cherenkov Telescopes (IACTs) for very high-energy (VHE) astronomy. The array shall be composed by three different types of telescopes, in order to maximize the performance in three different energy ranges: the Large Size Telescope (LST) for the low energy range (E $\simeq$ 20 GeV -- 1 TeV), the Medium Size Telescope (MST) for the core energy range (E $\simeq$ 0.1--10 TeV), and the Small Size Telescope (SST) for the high energy range (E $>$ 1 TeV). The ASTRI project (`Astrofisica con Specchi a Tecnologia Replicante Italiana') is included in this framework: it is a `flagship project' of the Italian Ministry of Education, University and Research, which, under the leadership of the Italian National Institute of Astrophysics (INAF), aims to realize and test an end-to-end prototype of the SST.
\vspace{-1cm}

\begin{figure}
\begin{center}
\psfig{file=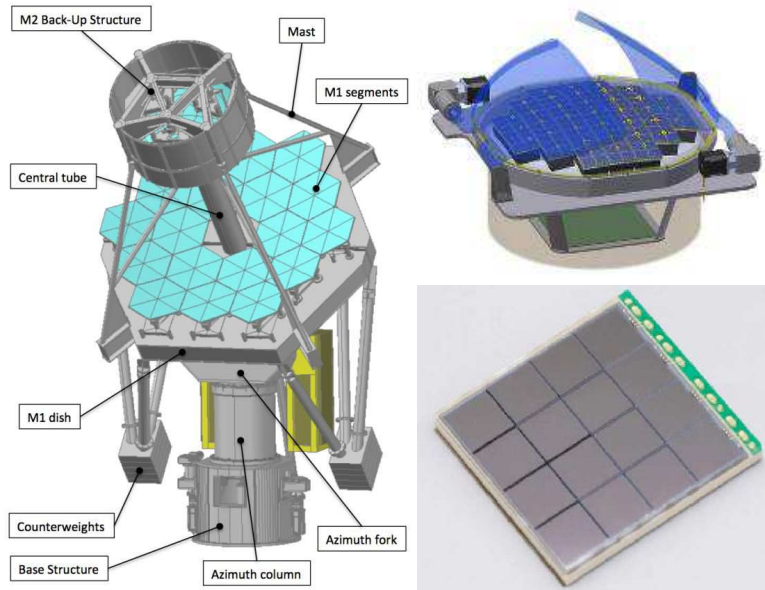,width=8.5cm,angle=-90}
\end{center}
\caption{Layout of the ASTRI SST-2M prototype (left panel), of its camera (upper right panel) and of one Silicon Photomultiplier (lower right panel).}
\label{fig1}
\end{figure}

\section{The ASTRI SST-2M prototype}\label{sec:prototype}

\smallskip
\noindent
\underline{{\bf Introduction.}} The ASTRI SST-2M prototype is characterized by two special features which will be adopted for the first time on a Cherenkov telescope (Pareschi et al. 2013\cite{Pareschi+13}): a dual-mirror Schwarzschild--Couder (SC) optical design (Vassiliev et al. 2007\cite{Vassiliev+07}), which is characterized by a wide field of view (FoV) and a compact optical configuration, and a light and compact camera based on Silicon photo-multipliers, which offer high photon detection sensitivity and fast temporal response. Figure~\ref{fig1} (left panel) shows the telescope layout, whose mount exploits the classical altazimuthal configuration.


\smallskip
\noindent
\underline{{\bf The Optical Design.}} The proposed layout (Canestrari et al. 2013\cite{Canestrari+13}) is characterized by a wide-field aplanatic optical configuration: it is composed by a segmented primary mirror made of three different types of segments, a concave secondary mirror, and a convex focal surface. The design has been optimized in order to ensure, over the entire FoV, a light concentration higher than 80 \% within the angular size of the pixels. The telescope design is compact, since the primary mirror (M1) and the secondary mirror (M2) have a diameter of 4.3 m and 1.8 m, respectively, and the primary-to-secondary distance is 3 m. The SC optical design has an f-number f/0.5, a plate scale of 37.5 mm/$^\circ$, a logical pixel size of approximately 0.17$^\circ$, an equivalent focal length of 2150 mm and a FoV of 9.6$^\circ$ in diameter; the mean value of the active area is $\sim$ 6.5 m$^2$.


\smallskip
\noindent
\underline{{\bf The Mirrors.}} The primary mirror is composed by 18 hexagonal segments, with an aperture of 849 mm face-to-face; the central segment is not used because it is completely obstructed by the secondary mirror. According to their distance from the optical axis, there are three different types of segments, each having a specific surface profile. In order to perform the correction of the tilt misplacements, each segment will be equipped with a triangular frame with two actuators and one fixed point. The secondary mirror is monolithic and has a curvature radius of 2200 mm and a diameter of 1800 mm. It will be equipped with three actuators, where the third actuator will provide the piston/focus adjustment for the entire optical system. For both the segments of the primary mirror and the secondary mirror the reflecting surface is obtained with a Vapor Deposition of a multilayer of pure dielectric material (Bonnoli et al. 2013\cite{Bonnoli+13}).


\smallskip
\noindent
\underline{{\bf The Camera.}} The SC optical configuration allows us designing a compact and light camera. In fact, the camera of the ASTRI SST-2M prototype has a dimension of about 56 cm $\times$ 56 cm $\times$ 56 cm, including the mechanics and the interface with the telescope structure, for a total weight of $\sim$ 50 kg (Catalano et al. 2013\cite{Catalano+13}). Such small detection surface, in turn, requires a spatial segmentation of a few square millimeters to be compliant with the imaging resolving angular size. In addition, the light sensor shall offer a high photon detection sensitivity in the wavelength range between 300 and 700 nm and a fast temporal response. In order to be compliant with these requirements, we selected the Hamamatsu Silicon Photomultiplier (SiPM) S11828-3344M. The `unit' provided by the manufacturer is the physical aggregation of 4 $\times$ 4 pixels (3 mm $\times$ 3 mm each pixel), while the logical aggregation of 2 $\times$ 2 pixels is a `logical pixel' (Figure~\ref{fig1}, lower right); its size of 6.2 mm $\times$ 6.2 mm corresponds to 0.17$^\circ$. In order to cover the full FoV, we adopt a modular approach: we aggregate 4 $\times$ 4 units in a Photon Detection Module (PDM) and, then, use 37 PDMs to cover the full FOV. The advantage of this design is that each PDM is physically independent of the others, allowing maintenance of small portions of the camera. To fit the curvature of the focal surface, each PDM is appropriately tilted with respect to the optical axis. The camera is also equipped with a light-tight two-petal lid (Figure~\ref{fig1}, upper right) in order to prevent accidental sunlight exposure of its SiPM detectors.

\smallskip
\noindent
\underline{{\bf The ASTRI SST-2M site.}} The ASTRI SST-2M prototype will be placed at the `M. G. Fracastoro Mountain Station', the observing site of the INAF Catania Astrophysical Observatory; it is at Serra La Nave, on the Etna Mountain, at an altitude of 1735 m a.s.l. (Maccarone et al. 2013\cite{Maccarone+13}). The prototype is currently under construction and it will be tested on field: it is scheduled to start data acquisition in 2014.

\smallskip
\noindent
\underline{{\bf The Prototype Expected Performance.}} Although the ASTRI SST-2M prototype will mainly be a technological demonstrator, it should be able to perform also scientific observations. Based on the foreseen sensitivity ($\simeq$ 0.2 Crab unit at 0.8 TeV), a source flux of 1 Crab at E $>$ 2 TeV should be detectable at 5 $\sigma$ confidence level in some hours, while a few tens of hours should be necessary to obtain a comparable detection at E $>$ 10 TeV (Bigongiari et al. 2013\cite{Bigongiari+13}). In this way we would obtain the first Crab observations with a Cherenkov telescope adopting a Schwarschild--Couder optical design and a SiPM camera; in addition, also the brightest AGNs (MKN 421 and MKN 501) could be detected.


\section{The ASTRI/CTA mini-array}
\vspace{-0.25cm}

\smallskip
\noindent
\underline{{\bf Introduction.}} Beside the prototype, the ASTRI project aims to realize, in collaboration with CTA international partners, a mini-array of a few SST dual-mirror (SST-2M) telescopes. Thanks to the array approach, it will be possible to check the trigger algorithms and the wide FoV performance, to compare the mini-array performance with the Monte Carlo expectations and to validate the performance predictions for the full SST array. The ASTRI/CTA mini-array shall constitute the first seed of the CTA Observatory at its Southern site and shall perform the first CTA science, starting operation in 2016.

\smallskip
\noindent
\underline{{\bf Performance.}} Considering 7 telescopes at an optimized distance of 250-300 m, preliminary Monte Carlo simulations yield a minimal improvement in sensitivity compared to the current IACTs (Di Pierro et al. 2013\cite{DiPierro+13}): starting from E = 10 TeV, the mini-array expected sensitivity is slightly better at few tens of TeV; moreover, the ASTRI/CTA mini-array sensitivity is still competitive up to 100 TeV, where the performance of the present generation of IACTs drops dramatically. In this energy regime the mini-array will operate as the most sensitive IACT array. The ASTRI/CTA mini-array will be able to study in great detail sources with a flux of a few 10$^{-12}$ erg cm$^{-2}$ s$^{-1}$ at 10 TeV, with an angular resolution of a few arcmin and an energy resolution of about 10-15 \%.

\smallskip
\noindent
\underline{{\bf Science.}} The ASTRI/CTA mini-array will observe prominent sources such as extreme blazars (1ES 0229+200), nearby well-known BL Lac objects (MKN 421 and MKN 501) and radio-galaxies, galactic pulsar wind nebulae (Crab Nebula, Vela-X, HESS 1825-137), supernovae remnants (Vela-junior, RX J1713.7-3946, Kepler) and microquasars (LS 5039), as well as the Galactic Center. In this way it will be possible to investigate the electron acceleration and cooling, to study the relativistic and non relativistic shocks, to search for cosmic-ray (CR) Pevatrons, to study the CR propagation and the impact of the extragalactic background light on the spectra of the nearby sources (Vercellone et al. 2013\cite{Vercellone+13}).
\vspace{-0.5cm}

\section*{Acknowledgments}

This work was partially supported by the ASTRI Flagship Project financed by the Italian Ministry of Education, University, and Research (MIUR) and lead by the Italian National Institute of Astrophysics (INAF). We also acknowledge partial support by the MIUR Bando PRIN 2009.

\end{document}